\documentstyle[12pt]{article}
\begin{document}
\begin{titlepage}
\begin{flushright}   CINVESTAV-FIS/96-12
		     \\
                     ICN-UNAM/96-22
\end{flushright}
\begin{center}
   \vskip 4em
  {\LARGE EXTENDED OBJECTS WITH EDGES}
  \vskip 5em
  {\large Riccardo Capovilla${}^{(1)}$ and
   Jemal Guven${}^{(2)}$\\[3em]}
\em{
 ${}^{(1)}$ Departamento de F\'{\i}sica \\
 Centro de Investigacion
y de Estudios Avanzados del I.P.N. \\
Apdo Postal 14-740, 07000 M\'exico,
D. F.,
MEXICO \\
}
\tt {capo@fis.cinvestav.mx} \\[1em]
\em{
${}^{(2)}$ Instituto de Ciencias Nucleares \\
 Universidad Nacional
Aut\'onoma de M\'exico\\
 Apdo. Postal 70-543, 04510 M\'exico, D.F., MEXICO \\
 }
\tt{
 jemal@nuclecu.unam.mx
}
\\[1em]
\end{center}
\vskip 3em
\begin{abstract}
We examine, from a geometrical point of view, the dynamics 
of a relativistic extended object with loaded edges. In the 
case of a Dirac-Nambu-Goto [DNG] object with  DNG edges, the 
worldsheet $m$ generated by the parent object is, as in the case 
without boundary, an extremal timelike surface in spacetime. Using 
simple variational arguments, we demonstrate that the worldsheet of each
edge is a constant mean curvature embedded timelike hypersurface on $m$, 
which coincides with its boundary, $\partial m$. The constant is 
equal in magnitude to the ratio of the bulk to the edge tension. The edge, 
in turn, exerts a dynamical influence on the motion of the parent
through the boundary conditions induced on $m$, specifically that the 
traces of the projections of the extrinsic curvatures of $m$  onto
$\partial m$ vanish. 
\end{abstract}
\date{\today}
\vskip 1em
PACS: 11.27.+d

\end{titlepage}
\newpage
\noindent{\bf I. INTRODUCTION}
\vskip 0.5cm

The lowest order phenomenological action describing  
the dynamics of a relativistic extended 
object, or membrane, is proportional to the area of its
worldsheet, $m$, and is known as the Dirac-Nambu-Goto [DNG] action
(For relevant examples, see \cite{VS,GSW,DPS} and, in the 
related statistical mechanical context, \cite{NPW}). 
The corresponding equations of motion of a closed 
object (without boundary) are completely described by the 
worldsheet diffeomorphism covariant system of non-linear
second order hyperbolic partial differential equations: 

\begin{equation}
K^i =0\,.\label{eq:k0}
\end{equation}
Here $K^i$ is the trace of the $i^{\rm th}$ extrinsic 
curvature of $m$ embedded 
in spacetime, one for each co-dimension of the embedding. 
In particular, the classical dynamics is entirely
independent of the tension of the membrane, $\mu_0$.. 

In this paper we focus on the modification required to 
this geometrical description when massive edges are admitted. Such edges may consist of several disconnected components.
Concrete examples consist of a segment of string with 
monopoles attached to its ends 
(a disconnected boundary) or a domain wall 
bounded by a string. The former is relevant in hadron physics as
an effective description of color flux tubes in QCD\cite{Nest}. 
The dynamics of such systems is also 
relevant in cosmology because objects of this type could have been 
generated if the early universe underwent an appropriate sequence
of phase transitions\cite{VS,Kibble}.

The key observation is that each edge
worldsheet is itself an embedded hypersurface in the worldsheet of
the parent membrane, which coincides with the boundary of the parent
worldsheet. The edges are thus treated as
membranes themselves, one dimension lower than the
parent membrane. 
The parent worldsheet is the spacetime where the edges live.
Since the parent membrane has a dynamics of its own, however,
this is no longer a fixed, prescribed background spacetime.
In the lowest order approximation, the  edges
will also  be described by a DNG action of the 
appropriate dimension with its own characteristic tension, 
$\mu_b$  (or mass, if pointlike).  The edge
worldsheet $\partial m$ will then satisfy\cite{RMF}

\begin{equation}
\mu_b k = - \mu_0 \,,\label{eq:k1}
\end{equation}
where $k$ is the trace of the extrinsic 
curvature of $\partial m$ embedded in $m$. 

There are two tractable approximations. 
In the limit that the mass of the edge tends to zero the 
null boundary dynamics associated with the theory of open 
membranes is recovered\cite{GSW}. 
This limit is the one adopted for
the open string action of string theory
in its most ambitious form as a theory of everything.
On the other hand, in the limit that the edge tension
goes to infinity,  $\mu_b\to\infty$, the edges 
themselves become extremal surfaces
of the background spacetime, and the membrane 
interpolates accordingly.
In particular, if the background spacetime
is flat, the edges can be assumed fixed. 
This approximation is frequently exploited
in the string approximation of the inter quark potential\cite{Nest}. 

The equations of motion (\ref{eq:k0}) and (\ref{eq:k1}) 
are not complete as they stand. What is missing is a statement 
about the dynamical feedback that the edges have on
the parent object spanning them. (This simple fact was overlooked
in Ref. \cite{RMF}.)
This is implemented in the form of constraints
on the extrinsic geometry of $m$ at the boundary. 
Specifically, one obtains that,

\begin{equation}
{\cal H}^{ab} K_{ab}^i =0\,,\label{eq:k00}
\end{equation}
where $K^i_{ab}$ is the $i^{\rm th}$ extrinsic 
curvature of $m$ embedded in spacetime, 
${\cal H}^{ab}$ is the projection operator in $m$ onto 
$\partial m$, and this equation is to be evaluated on
$\partial m$. These constraints 
must be implemented as boundary conditions on Eqs.(\ref{eq:k0}).
Suppose we had failed to implement these conditions.
Then any given initial conditions on the 
boundary which are tangent to the worldsheet of 
a closed solution to Eqs.(\ref{eq:k0}) would simply generate a
timelike hypersurface on this worldsheet. 
The only feedback it would have on 
the parent membrane would be to determine 
the limits of the truncation of the closed solution. 
On the contrary, the boundary conditions (\ref{eq:k00}) generally
place stringent conditions on the motion 
of the membrane. They are, however, vacuous when the
membrane is totally geodesic, $K_{ab}^i=0$.  Such is the case
of a planar worldsheet describing, for example,
a non-rotating string or a planar disc of membrane in Minkowski
space. 

For the case of a string with massive ends, the boundary
equation of motion (\ref{eq:k1}) and boundary
conditions (\ref{eq:k00}) can be 
cast in a particularly attractive form.
If the trajectory of an end 
is parametrized  with proper time $\tau$, 
then Eqs. (\ref{eq:k1}) and (\ref{eq:k00}) can be 
combined to give:

\begin{equation}  
\mu_b D_\tau \left({d X^\mu \over d\tau}\right) 
= - \mu_0 \,\eta^\mu \,,
\end{equation}
where $\eta^\mu$ is the inward normal to the boundary of the string
worldsheet, and  $D_\tau :=(d X^\mu/d\tau)D_\mu$ is the
projection onto the end worldline of the spacetime covariant
derivative. 
The acceleration of an end  is therefore of constant magnitude
and directed into the string worldsheet.

This paper is organized as follows:
In sect.II we provide a summary of the relevant mathematical
formalism. In particular, we discuss the connection between the 
hierarchy of embeddings: $\partial m$ in $m$, 
$m$ in spacetime, and the direct embedding of the edges in spacetime. 
In order to simplify our presentation we confine our attention
to the case of an extremal membrane, described by
the DNG action, with edges described 
by a DNG action of one lower dimension.
In sect.III we derive the complete equations of motion for this system, 
Eqs.(\ref{eq:k0}), (\ref{eq:k1}) and (\ref{eq:k00}). 
What is remarkable is just how efficient variational 
principles are in isolating the appropriate boundary conditions\cite{Lanczos}. 
We conclude in Sect. IV with a brief discussion that focuses
on a rotating string with massive ends.

\vskip 1cm
\noindent{\bf II. KINEMATICS}
\vskip 0.5cm
To begin with, consider an oriented {\it timelike}
worldsheet $m$ of dimension $D$,
which corresponds to the trajectory of the membrane
in  an $N$-dimensional spacetime $\{ M , g_{\mu\nu} \}$. 
The worldsheet $m$ is described by the embedding\cite{MATH}  

\begin{equation}
x^\mu=X^\mu(\xi^a)\,,\label{eq:1}
\end{equation}
where $x^\mu$ are coordinates on $M$, and $\xi^a$ coordinates
on $m$ ($\mu, \nu, \cdots = 0, \cdots, N-1$, and $a, b, \cdots = 0, 
\cdots, D-1$). 
The $D$ vectors,
\begin{equation}
e_a :=X^\mu_{,a}\partial_\mu\,, \label{eq:2}
\end{equation}
form a basis of tangent vectors to 
$m$, at each point of $m$.
The Lorentzian metric induced on the worldsheet is then 
given by,
\begin{equation}
\gamma_{ab}= e^\mu{}_a e^\nu{}_b\,g_{\mu\nu} 
\,.\label{eq:3}
\end{equation}
Note that in statistical mechanics applications 
we are interested in a Euclidean ``spatial" 
metric $g_{\mu\nu}$, and, of course, the
induced metric $\gamma_{ab}$ is Euclidean as well.

Let the spacetime vectors $n^{\mu\, i}$ denote 
the $i^{\rm th}$ unit normal to the worldsheet 
($i, j, \cdots = 1, \cdots, N-D$), defined, up to  
a local $O(N-D)$ rotation, with
\begin{equation}
g_{\mu \nu} e^\mu{}_a \, n^{\nu\,i} = 0\,, \; \; \; 
g_{\mu \nu} n^{\mu \, i} n^{\nu \, j} = \delta^{ij}\,.
\label{eq:4}
\end{equation}
Normal indices are raised and 
lowered with $\delta^{ij}$ and $\delta_{ij}$, respectively, whereas
tangential indices are raised and lowered with $\gamma^{ab}$ and 
$\gamma_{ab}$, respectively. 

The vectors $\{ e_a , n_i \}$ form a basis for spacetime 
vectors adapted to the situation of interest here.

The worldsheet projection of the 
spacetime covariant derivatives is defined by $D_a := e_a^\mu D_\mu$,
where $D_\mu$ is the (torsionless) covariant derivative compatible
with $g_{\mu \nu}$.
The classical Gauss-Weingarten equations (see \cite{MATH,CG1}) are given by, 

\begin{eqnarray}
D_a e^\mu{}_b &=& \gamma_{ab}{}^c e^\mu{}_c - K_{ab}{}^i n^\mu{}_i\,,  
\label{eq:gauss} \\
D_a n^\mu{}_i &=& K_{ab\,i} e^{\mu b } + \omega_{a\,i}{}^j n^\mu{}_j\,.
\label{eq:wein}
\end{eqnarray}
The $\gamma_{ab}{}^c = \gamma_{ba}{}^c$ are the 
connection coefficients compatible with the worldsheet metric
$\gamma_{ab}$. The quantity $K_{ab}{}^i$ is the 
$i^{\rm th}$ extrinsic curvature of the worldsheet defined by

\begin{equation}
 K_{ab}{}^i = - g_{\mu \nu} n^{\mu\, i} D_a e^\nu{}_b  
 = K_{ba}{}^i\,. \label{eq:6} 
\end{equation}
The extrinsic geometry of $m$ is determined by $K_{ab}{}^i$, 
and by the extrinsic twist potential, $\omega_{a\,ij}$
associated with the covariance under normal frame rotations. 
(see {\it e.g.} \cite{JG,CG1}).

Not every specification of the 
intrinsic and of the extrinsic geometry is 
necessarily consistent with some embedding.
There are integrability conditions, the
Gauss-Codazzi, Codazzi-Mainardi, and Ricci equations,
which must be satisfied by the intrinsic
and extrinsic geometry, for an embedding to exist.
We will return to these equations below in 
the context of the boundary.

We turn now to the definition of the intrinsic and extrinsic geometry
of the worldsheet boundary $\partial m$. We 
treat $\partial m$ as a timelike surface of dimension
$D-1$, described by the embedding in the worldsheet $m$,

\begin{equation}
\xi^a = \chi^a ( u^A) \label{eq:7}
\end{equation}
where $A,B,... = 0, 1, \cdots , D-2$, and $u^A$ are coordinates on
$\partial m$.

The definition of the extrinsic and intrinsic geometry 
of the worldsheet boundary provides a special case of the
discussion given above for an arbitrary worldsheet.
In order to establish our notation, we repeat it,
specializing  to the case of co-dimension one.
The $D-2$ vectors
$ \epsilon_A := \chi^a{}_{,A} \partial_a$
are tangent to the boundary worldsheet $\partial m$.
The metric induced on $\partial m$ is then,

\begin{equation}
 h_{AB} = \gamma_{ab} \chi^a{}_{,A} \chi^b{}_{,B}\,. 
\label{eq:8}
\end{equation}
The normal to $\partial m$ is defined by 

\begin{equation}
 \gamma_{ab} \eta^a \epsilon^b{}_A = 0 , \;\;\;\;\;\; 
\gamma_{ab} \eta^a \eta^b  = 1\,. \label{eq:bn}
\end{equation}
The Gauss-Weingarten equations take the form:

\begin{eqnarray}
 \nabla_A \epsilon^a{}_B &=& \gamma_{AB}{}^C \epsilon^a{}_C
 - k_{AB} \eta^a\,, \label{eq:gauss1}\\
\nabla_A \eta^a &=& k_{AB} \epsilon^{a\,B} \,.
\label{eq:wein1}
\end{eqnarray}
where $ \nabla_A = \epsilon^a{}_A \nabla_a $ is the gradient along
the tangential basis $\{ \epsilon^A \}$, $\gamma_{AB}{}^C $ are the
connection coefficients compatible with the boundary worldsheet metric
$h_{AB}$, and $k_{AB} = k_{BA}$ is the edge worldsheet
extrinsic curvature associated with the embedding of
$\delta m$ in $m$. For a co-dimension one embedding, the extrinsic 
geometry is determined completely by the extrinsic curvature,
and the Ricci integrability conditions are vacuous.

For the role it will play in the sequel,
it is useful also to contrast this description with 
the description of the boundary $\delta m$, or, which is the same thing,
of the edge worldsheet, embedded directly in spacetime, 

\begin{equation}
x^\mu = X^\mu( u^A)\,,
\end{equation}
with tangents $e_A^\mu:= e_a^\mu \epsilon^a_A$. This corresponds
to the map composition $X^\mu (\xi^a ( u^A)) = X^\mu( u^A)$.
The induced metric is exactly as before, Eq. (\ref{eq:8}). 
The spacetime normals to $m$
are also normal to $\partial m$ in spacetime. With $\eta^\mu
:=e^\mu_a\eta^a$, these vectors complete the normal basis which we 
label $n^I:=\{\eta, n^i\}$. We will use the index $0$ to denote
the direction along $\eta^\mu$. It should not be confused with a
timelike index. We can now write down the 
corresponding Gauss-Weingarten equations ($D_A:=e^\mu_A D_\mu$),

\begin{eqnarray}
D_A e^\mu{}_B &=& \gamma_{AB}{}^C e^\mu{}_C - K_{AB}{}^I n^\mu{}_I\,,  
\label{eq:gauss2} \\
D_A n^\mu{}_I &=& K_{AB\,I} e^{\mu\,B} + \omega_{A\,I}{}^J n^\mu{}_J\,.
\label{eq:wein2}
\end{eqnarray}
With respect to this adapted basis, it is simple to check that
$K^i_{AB}= \epsilon^a_A\epsilon^b_B K^i_{ab}$, and 
$K^0_{AB}= k_{AB}$. In addition, $\omega_{A\,ij} = \epsilon^a_A
\omega_{a\,ij}$. The boundary inherits the 
extrinsic curvature and twist of the worldsheet. However, note that 
$\omega_{A\,i0} =\eta^a \epsilon^b_A K_{ab\,i}$. Thus, there is the 
possibility that the boundary worldsheet might have a non 
trivial twist (associated with its embedding in spacetime)
though the parent worldsheet does not.
In particular, this might be the case when the parent worldsheet is embedded as a 
hypersurface in spacetime. In the case of a one-dimensional boundary,
however, the extrinsic twist will be pure gauge.

In is also instructive to examine the hierarchy of integrability 
conditions which emerges in these alternative embeddings of the 
boundary. On one hand, we have the Gauss-Codazzi, 
Codazzi-Mainardi, and Ricci integrability conditions associated with
the embedding of $m$ in spacetime:

\begin{equation}
R_{abcd} = {\cal R}_{abcd} -
K_{ac}{}^{i} K_{bd\,i} + K_{ad}{}^i K_{bc\,i}\,, 
\label{eq:gc1}
\end{equation}

\begin{equation}
R_{abc\,i} = \widetilde\nabla_a K_{bc\,i} - 
\widetilde\nabla_b K_{ac\,i}\,,
\label{eq:cm1}
\end{equation}
and

\begin{equation}
R_{ab\,ij} = 
 \Omega_{ab\, ij} -
K_{ac\,i} K_b{}^{c}{}_j
+ K_{bc\,i} K_a{}^{c}{}_j\,.
\label{eq:ricci1}
\end{equation}
Here $\widetilde\nabla_a$ is the covariant derivative 
associated with the extrinsic twist potential $\omega_a{}^{ij}$, and
$\Omega_{ab\,ij}$ is its curvature. The left hand side of these
equations denote the contraction of the background spacetime 
Riemann tensor $R^\mu{}_{\nu \rho \sigma}$ with the basis $\{ e_a , n^i \}$.
${\cal R}^a{}_{bcd}$ is the Riemann tensor of the worldsheet 
covariant derivative $\nabla_a$.

We also have the Gauss-Codazzi and 
Codazzi-Mainardi integrability conditions associated with
the embedding of $\partial m$ in $m$:

\begin{equation}
{\cal R}_{ABCD} = {\rm R}_{ABCD} -
k_{AC} k_{BD} + k_{AD} k_{BC}\,, 
\end{equation}
and

\begin{equation}
{\cal R}_{ABCd}\eta^d 
= {\cal D}_A k_{BC} - 
{\cal D}_B k_{AC}\,.
\end{equation}
The left hand side of these
equations denote the contraction of the worldsheet 
Riemann tensor ${\cal R}^a{}_{bcd}$ with the basis $\{ \epsilon_A , \eta \}$.
We use the notation ${\rm R}_{ABCD}$ 
to denote the Riemann tensor of the boundary  
covariant derivative ${\cal D}_A$.

Finally, there are the Gauss-Codazzi, 
Codazzi-Mainardi, and Ricci integrability conditions associated with the 
direct embedding of the boundary in spacetime:

\begin{equation}
R_{ABCD} 
= {\rm R}_{ABCD} -
K_{AC}{}^{I} K_{BD\,I} + K_{AD}{}^I K_{BC\,I}\,, 
\end{equation}

\begin{equation}
R_{ABCI}
= \widetilde{\cal D}_A K_{BC\,I} - 
\widetilde{\cal D}_B K_{AC\,I}\,,
\end{equation}
and

\begin{equation}
R_{ABIJ}
=  \Omega_{AB\, IJ} -
K_{AC\,I} K_B{}^{C}{}_J
+ K_{BC\,I} K_A{}^{C}{}_J\,.
\label{eq:ricci3}
\end{equation}
$\widetilde{\cal D}_A$ is the twist covariant derivative associated
with $\omega_A{}^{IJ}$.
We note that  consistency between Eq.(\ref{eq:ricci1}) and
(\ref{eq:ricci3}) implies

\begin{eqnarray}
\Omega_{AB\, ij} &=&\epsilon ^a_A\epsilon^b_B \Omega_{ab\, ij}\,, 
\nonumber\\
 \Omega_{AB\, i0} &=&
\epsilon^c_C \left[\epsilon^a_A  K_{ac\,i} k_B{}^{C}
-\epsilon^b_B K_{bc\,i} k_A{}^{C}\right] \,. \nonumber
\end{eqnarray}

\vskip 1cm
\noindent{\bf III. EXTREMAL OBJECTS WITH LOADED EDGES}
\vskip 0.5cm

The dynamics of the membrane is specified by the choice of
an appropriate phenomenological action, constructed with
scalars built with the quantities that characterize
the intrinsic and extrinsic geometry of the
membrane worldsheet. In the presence of edges, one needs also
to specify  some dynamical rule for the edges
themselves. We choose the DNG 
action for the membrane, and the same action for its edges.

The action we consider is

\begin{equation}
 S = S_{0} + S_b\,, \label{eq:nambu}
\end{equation}
where

\begin{eqnarray}
  S_{0} [ X,\chi ] &=& -\mu_{0} \int_m d^D \xi 
\sqrt{- \gamma }\,, \label{eq:12a} \\
S_b [\chi,X] &= &- \mu_b \int_{\partial m} d^{D-1} u
\sqrt{ - h}\,, 
\label{eq:12b}
\end{eqnarray}
$\mu_{0}$ is the membrane tension, 
$\mu_b$ is the tension of the  edge membrane,
$\gamma$ the determinant of the
membrane worldsheet metric $\gamma_{ab}$, and $h$
is the determinant of the boundary worldsheet metric $h_{AB}$.
This action is a functional both of the embedding $X^\mu$ of $m$ in $M$,
and of the embedding $\chi^a$ of $\partial m$ in $m$.
There may well be many disconnected edges. To avoid clutter
we ascribe the one tension $\mu_b$ to all.

To derive the equations of motion arising from the action (\ref{eq:nambu}), 
consider first a variation of the embedding of $m$, 
$X^\mu \rightarrow X^\mu + 
\delta X^\mu$. The displacement is assumed to 
vanish on two spacelike hypersurfaces of $m$, which play
the role of initial and final times.

We decompose the displacement with respect to the
spacetime basis $\{ e^a , n^i \}$, as

\begin{equation}
\delta X = \Phi^a e_a + \Phi^i n_i\,. \label{eq:delta}
\end{equation}
We now have that under this displacement, the intrinsic metric
change is \cite{CG1,ALG} 

\begin{equation}
\delta_X \gamma_{ab} = 2 K_{ab}^i\Phi_i 
+ \nabla_a\Phi_b + \nabla_b\Phi_a \,.\label{eq:delgam}
\end{equation}
The variation of the membrane action $S_{0}$ gives,

\begin{eqnarray}
 \delta_X S_{0} 
&=& -  \mu_{0}
\int_m d^D \xi\, 
\sqrt{\gamma}\gamma^{ab} \left[
K_{ab}^i\Phi_i 
+ \nabla_{a}\Phi_{b} \right]\nonumber\\
&=& -  \mu_{0}
\int_m d^D \xi\, 
\sqrt{\gamma}\left[  K^i\Phi_i
+ \nabla_a\Phi^a\right]\\
&=& - \mu_{0}
\int_m d^D \xi \sqrt{- \gamma} 
K^i \Phi_i
-  \mu_{0}
\int_{\partial m} d^{D-1} u  \sqrt{- h}\, \eta^b \Phi_b\,. 
\label{eq:16}
\end{eqnarray}
The last line obtains from the preceeding one by applying
Stokes' theorem to the second term. Here $\eta^a$ is the outward
pointing normal to $\partial m$ introduced in 
Eq.(\ref{eq:bn}). 
We find that only the normal projection of the variation, $\Phi_i$,
contributes to the equations of motion of the membrane,  
this is generally true regardless of the form of the action $S_0$
so long as it is constructed in a 
worldsheet ($m$) diffeomophism invariant way.
There is no boundary term associated with $\Phi_i$. 
This is not, however, generally true,
it is an artifact of extremal dynamics.

The tangential variation gives only a boundary term. This is 
a consequence of the fact that tangential deformations correspond
modulo a displacement of the boundary to 
infinitesimal worldsheet diffeomorphisms. This is why 
we could ignore such variations in our study of objects without boundary. 

We note that the boundary contribution to 
Eq.(\ref{eq:16}) is (minus $\mu_0$ times) the
change in the worldsheet volume under a normal deformation of the 
boundary, $\delta \chi^a = \eta^b \Phi_b \eta^a$. 
As we will see, it will contribute to the equations of motion of the edge. 
The projections of $\Phi^b$ onto $\partial m$ do not contribute.

Before considering the variational analysis of $S_b$, 
we comment briefly on the case of an open membrane with a massless boundary.
The normal term will vanish whenever
$\gamma(\eta,\eta)=0$ or the boundary is null.
Physically, no momentum may cross the surface.
This can only occur if the boundary is a null surface, moving at each point 
at the speed of light [2]. In the textbook treatment this is arranged 
by demanding that the normal projection of the 
worldsheet derivative of the embedding function
vanish on the boundary. (For a geometric treatment of such boundary
conditions, see \cite{HT})

Let us consider now the variation of the edge action
under the infinitesimal variation in the worldsheet (\ref{eq:delta}).
This variation is transmitted to the geometry of the boundary through
its effect on $\gamma_{ab}$, given by Eq.(\ref{eq:delgam}).
We have

\begin{eqnarray}
 \delta_X S_b &=& - {1\over 2} \mu_{b} \int_{\partial m }
d^{D-1} u \sqrt{-h} h^{AB}\delta_X h_{AB}\nonumber\\
&=& - \mu_{b} \int_{\partial m }
d^{D-1} u \sqrt{-h} {\cal H}^{ab}
(K_{ab}^i \Phi_i + \nabla_{a}\Phi_{b}) \,, \label{eq:17}
\end{eqnarray}
where we have introduced the projector onto
$\partial m$, ${\cal H}^{ab}:=h^{AB}\chi^a_{,A}\chi^b_{,B}$
and the fact that $\delta_X h_{AB} = 
\chi^a_{,A}\chi^b_{,B} \delta_X \gamma_{ab}$.

The vanishing of the variation of the total action $S$
(\ref{eq:nambu})
under arbitrary normal deformations, $\Phi_i$,
gives the equations of motion for the membrane, Eq. (\ref{eq:k0}), $K^i = 0$. 
$\Phi_i$ is not fixed on the boundary, so there is a boundary term 
appearing in Eq.({\ref{eq:17}) to contend with. 
It vanishes whenever , Eq.(\ref{eq:k00}),
${\cal H}^{ab} K_{ab}{}^i = 0$, evaluated on $\partial m$, is satisfied.
The variational principle
has therefore also provided the natural boundary conditions 
on the embedding $X$. We will discuss the interpretation 
of these conditions below.

The vanishing of the variation of $S$, 
under arbitrary tangential deformations, $\Phi_a$
with support on the boundary,
gives the equation of motion for the boundary (\ref{eq:k1}),(if $\mu_b\ne 0$)
$ k = - \mu_0 / \mu_b $.
To see this we note that

\begin{equation}
{\cal H}^{ab}\nabla_{a}\Phi_{b} 
= {\cal D}_A \Phi^A + k \eta^a\Phi_a \,,\label{eq:sum}
\end{equation}
where we have exploited the fact that 
$k = {\cal H}^{ab} \nabla_a \eta_b$ and 
we define $\Phi_A = \gamma_{ab}\Phi^a\epsilon^b_{A}$, so that

\begin{equation}
\delta_X S_b = - \mu_{b} \int_{\partial m }
d^{D-1} u \sqrt{-h} ( {\cal H}^{ab} K_{ab}^i \Phi_i +  
{\cal D}_A \Phi^A + k \eta^a\Phi_a )\,.
\end{equation} 
The first term appearing on the right hand side of Eq.(\ref{eq:sum})
is a divergence --- corresponding
to an infinitesimal boundary diffeomorphism, 
$\delta\chi^A =\Phi^A$. 
In the case of a {\it smooth} physical boundary this 
term does not contribute (the boundary of a boundary is zero).
The latter term appearing on the right hand side of Eq.(\ref{eq:sum}), 
however,
adds to the surface term appearing in Eq.(\ref{eq:16})
to give Eq.(\ref{eq:k1}).

We have not had to vary the action with 
respect to the boundary embedding to obtain Eq.(\ref{eq:k1}).
For completeness, and consistency, let us now consider the variation in 
$S$ induced by a displacement of the boundary worldsheet 
$\partial m$,

\begin{equation}
 \delta \chi = \Psi \eta + \Psi^A \epsilon_{A}\,.
\label{eq:18} 
\end{equation}
We obtain 

\begin{equation}
 \delta_\chi S = 
-  \mu_{0}
\int_{\partial m} d^{D-1} u  \sqrt{- h}\, \Psi
-\mu_{b} \int_{\partial m }
d^{D-1} u \sqrt{-h} h^{AB} k_{AB} \Psi\,, \label{eq:18b}
\end{equation}
modulo the same divergence appearing in Eq.(\ref{eq:sum}).
We again reproduce Eq.(\ref{eq:k1}), nothing new is obtained.
It is worth noting that this variation does not pick up
Eq.(\ref{eq:k00}).

It is instructive to compare the equations of 
motion describing the dynamics of an isolated boundary (imagine 
the spanning membrane removed) with 
Eq.(\ref{eq:k1}). The dynamics is now simply extremal and we have
(in the notation of sect.II) $h^{AB} K^I_{AB}=0$, or 
alternatively $k=0$ and ${\cal H}^{ab} K^i_{ab} =0$. 
The former differs from (\ref{eq:k1}) in the manner we would expect. 
The latter set of equations, however, reproduce the 
boundary conditions given by Eqs.(\ref{eq:k00}).
The departure from extremality when the 
boundary is spanned by a membrane occurs along the normal
which is tangent to the membrane worldsheet. 

The boundary conditions (\ref{eq:k00}) 
are still not exactly the standard (Robin) kind of
boundary condition we are accustomed to handle. It is 
worthwhile therefore to demonstrate explicitly that
they are sensible boundary conditions on Eqs.(\ref{eq:k0}).
We note that 

\begin{equation}  
K_{ab}^i= - n^i_\mu \left(\nabla_a\nabla_b X^\mu 
+ \Gamma^\mu_{\alpha\beta} X^\alpha_{,a} X^\beta_{,b}\right)\,,
\end{equation}
so that Eq.(\ref{eq:k00}) reads

\begin{equation}  
n^i_\mu \Big[\left(\Delta
- \eta^a\eta^b\nabla_a\nabla_b\right) X^\mu 
+ \Gamma^\mu_{\alpha\beta} {\cal H}^{ab} X^\alpha_{,a} X^\beta_{,b}\Big]
=0\,.\label{eq:k00X}
\end{equation}
We now exploit the fact that the Laplacian,  
$\Delta \Psi$, of any worldsheet scalar (such as 
$X^\mu$) can be decomposed as

\begin{equation}
\Delta \Psi = {\cal D}^A {\cal D}_A \Psi + 
(\eta^a \nabla_a)^2 \Psi + k \eta^a \nabla_a \Psi\,,
\end{equation}
and the fact that

\begin{equation}
\eta^a\eta^b\nabla_a\nabla_b \Psi = (\eta^a \nabla_a)^2 \Psi \,,
\end{equation}
to express Eq.(\ref{eq:k00X}) in the alternative form

\begin{equation}  
n^i_\mu \Big[
{\cal D}^A {\cal D}_A  X^\mu 
+  \Gamma^\mu_{\alpha\beta} {\cal H}^{\alpha\beta}\Big]
=0\,.\label{eq:k00X1}
\end{equation}
where we have defined 
${\cal H}^{\alpha\beta} := {\cal H}^{ab} X^\alpha_{,a} X^\beta_{,b}$.
In this form, the equations 
Eq.(\ref{eq:k00}) involve only  
derivatives of $X^\mu$ along $\partial m$, and thus it provides
sensible boundary  conditions for Eq. (\ref{eq:k0}).

Moreover, the form (\ref{eq:k00X1}) of the boundary conditions
suggests to reexpress the edge equations of motion, Eq. (\ref{eq:k1}),
as

\begin{equation}  
\eta_\mu \Big[{\cal D}^A {\cal D}_A  X^\mu 
+  \Gamma^\mu_{\alpha\beta} {\cal H}^{\alpha\beta}\Big]
= - {\mu_0 \over \mu_b }\,,\label{eq:k1X}
\end{equation}
and we can now combine Eqs.(\ref{eq:k00X1}) and (\ref{eq:k1X}) 
as

\begin{equation}  
{\cal D}^A {\cal D}_A  X^\mu 
+  \Gamma^\mu_{\alpha\beta} {\cal H}^{\alpha\beta}
= - {\mu_0 \over \mu_b } \eta^\mu \,.\label{eq:k1XX}
\end{equation}
This equations exhibit clearly the effect of the spanning
membrane on the dynamics of the edges, via the driving
term on the right hand side.
In the case of a string, with proper time $\tau$ 
along the trajectory of a boundary point, Eq.(\ref{eq:k1XX}) reduces to
($D_\tau =(d X^\mu/d\tau)D_\mu$)

\begin{equation}  
\mu_b D_\tau \left({d X^\mu \over d\tau}\right) 
= - \mu_0 \,\eta^\mu \,,\label{eq:k1XXb}
\end{equation}
so that the acceleration of a boundary point
$D_\tau (d X^\mu/d\tau)$ is constant in magnitude and directed into
$m$.

\vskip 1cm
\noindent{\bf IV. DISCUSSION}
\vskip 0.5cm

Consider the example of a rigidly rotating string bounded by point particles. 
It is clear that there is no solution of Eq.(\ref{eq:k0}) 
corresponding to a straight 
{\it non-rotating} segment of string with massless ends. Energy conservation 
would imply that such a configuration has a fixed proper length which 
is inconsistent with the nullity of the ends. With masses 
loading the ends, however, a solution exists
because energy can be transfered from the string to its boundary.
The monopoles are accelerated towards 
each other by the constant force provided by the tension in the string, the string collapses to a singularity. 

When the string rotates, the massive ends experience a 
centrifugal acceleration. Our non-relativistic 
intuition suggests that stable bound states 
exist. In particular, circular orbits with 
a fixed radius, $R$ (corresponding
to a fixed string length) and fixed angular velocity $\omega$ exist. 
These orbits are constrained by the requirement that 
$\omega R \le 1$. The 
corresponding worldsheet of the string is simply a truncation
at this radius of the circular timelike helicoid 
of a rigidly rotating string with massless ends. 
Geometrically, this is possible because
the boundary conditions Eq.(\ref{eq:k00}) are automatically satisfied when $\omega$ and $R$ are constants.

In the higher dimensional case of a 
membrane bounded by a string, a non-trivial interplay between the 
tension in the membrane and that in the boundary is possible. 
These forces
might operate in opposite directions. This is the case for 
a circular hole in a planar sheet of membrane. The tension on 
the circle tends to restore the membrane, that in the membrane to 
self destruction. There is clearly a critical radius 
determining which one will prevail. This competition is expected to
play a role in topology changing processes. 

In a subsequent publication we examine perturbation theory
pointing out, in particular, how we must modify the 
treatment in \cite{JG} or \cite{CG1} when dynamical boundaries are taken into 
account\cite{B2}.
In \cite{B3} the analysis undertaken here for
DNG extended objects is generalized to arbitrary 
phenomenological actions, both for the membrane and for the boundary. 
\newpage

\noindent{\bf Acknowledgements}

\vspace{.3cm}

We gratefully acknowledge support from CONACyT grant no.
211085-5-0118PE.

\end{document}